\documentclass[12pt]{iopart}

\usepackage{graphicx}
\usepackage{amsfonts}

\usepackage{epsfig}
\usepackage{graphicx}
\usepackage{dcolumn}
\usepackage{bm}
\usepackage{times}
\usepackage{xcolor}
\usepackage{colortbl,booktabs}
\usepackage{makecell}
\usepackage{amstext}
\usepackage{amssymb} 

\usepackage{float}
\usepackage{verbatim}

\begin{document}

\title[Detecting network communities via greedy expanding based on local superiority index]{Detecting network communities via greedy expanding based on local superiority index
}    

\author{Junfang Zhu$^{1,2}$, Xuezao Ren$^{2*}$, Peijie Ma$^{2}$, Kun Gao$^{2}$, Bing-Hong Wang$^{3*}$ and Tao Zhou$^{1*}$} 

\address{$^1$Big Data Research Center, University of Electronic Science and Technology of China, Chengdu 611731, People's Republic of China}

\address{$^2$School of Science, Southwest University of Science and Technology, Mianyang 621010, People's Republic of China}

\address{$^3$Department of Modern Physics, University of Science and Technology of China, Hefei 230026, People’s Republic of China}

\eads{\mailto{renxuezao@aliyun.com}, \mailto{bhwang@ustc.edu.cn}, \mailto{zhutou@ustc.edu}}


\begin{abstract}

Community detection is a significant and challenging task in network science. Nowadays, plenty of attention has been paid on local methods for community detection. Greedy expanding is a popular and efficient class of local algorithms, which typically starts from some selected central nodes and expands those nodes to obtain provisional communities by optimizing a certain quality function. In this paper, we propose a novel index, called local superiority index (LSI), to identify central nodes. In the process of expansion, we apply the fitness function to estimate the quality of provisional communities and ensure that all provisional communities must be weak communities. Evaluation based on the normalized mutual information suggests: (1) LSI is superior to the global maximal degree index and the local maximal degree index on most considered networks; (2) The greedy algorithm based on LSI is better than the classical fast algorithm on most considered networks.

\end{abstract}

\maketitle

\section{Introduction}
Community structure \cite{Mucha2010} is widely observed in various types of real networks such as social networks \cite{Zachary1977}-\cite{Chouchani2020}, biological networks \cite{Jonsson2006}-\cite{Cui2020}, technological networks \cite{Newman2004a}, and so on. A community is a group of nodes with more edges connecting nodes within the group and less edges connecting nodes to different groups \cite{Girvan2002}. Community detection is of significant importance in understanding organization principles and predicting dynamical behaviors of a network \cite{Fortunato2010}, which has already found wide applications in interdisciplinary domains, such as to find the pathways between diseases and drugs \cite{Pham2019}, to reveal the role of each part in a layered neural network \cite{Watanabe2019}, to mine user opinions from social networks \cite{Li2019}, and so on.  

Since the past decades, many methods of community detection have been proposed, such as graph partitioning \cite{Shi2011}, spectral clustering \cite{Donetti2004}-\cite{Jin2015}, similarity expansion \cite{Xiang2009}-\cite{Pan2010}, statistical inference \cite{Newman2007}, dynamic methods \cite{Zhou2003}-\cite{Hu2008}, and integer programming model \cite{Srinivas2019}. Among them, greedy algorithms based on quality function optimization have attracted a lot of attention \cite{Newman2004b}-\cite{Shao2020}. 
In a typical greedy algorithm, some nodes are selected as central nodes of communities (others are non-central nodes), from which the nearest neighbor search is usually adopted to expand provisional communities \cite{Saha2016}. Obviously, the selection of central nodes are vital for the accuracy of a greedy algorithm. In an early work \cite{Chen2013}, nodes either with global maximal degrees (GMD) or with local maximal degrees (LMD) are regarded as central nodes. The GMD takes nodes of the top-$k$ highest degrees as central nodes. However, the selection of the value of $k$ is often arbitrary. On the other hand, nodes with maximal degrees are not always central nodes \cite{Blondel2008b,Lu2016}. In contrast, the LMD designates nodes with higher degree than all neighbors as central nodes. Empirical evidences \cite{Chen2013} suggest that the LMD performs overall better than the GMD in identifying central nodes of provisional communities. 

In this paper, we propose a novel local centrality index called the local superiority index (LSI) and combine it to the greedy algorithm to detect communities. In our proposed algorithm, the community expansion, aiming to maximize the value of the quality function, starts from central nodes selected by the LSI. After expansion, we allot residual nodes to suitable communities. Allocation of residual nodes may decrease the quality function, while it is found that the optimal community structure is not always with the largest quality function \cite{Zhang2014}. Our algorithm is evaluated by normalized mutual information (NMI) \cite{Kuncheva2004} on real networks with known communities and benchmark networks generated artificially \cite{Lancichinetti2008}.

\section{Methods}

\begin{figure}[H]
\setlength{\abovecaptionskip}{0.cm}
\setlength{\belowcaptionskip}{-0.cm}
\centering
	\includegraphics[width=0.9\textwidth]{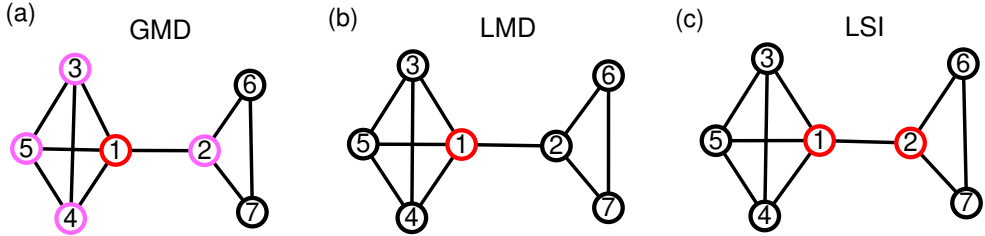}
    \caption{Comparison among the three considered indices. The illustrated network is constructed by connecting two complete networks through a single edge. (a) According to the GMD, the red node (node $1$) is the top-$1$ highest-degree node, the pink nodes (nodes $2,3,4,5$) and node $1$ are the top-$5$ highest-degree nodes. (b) According to the LMD, only node $1$ is a central node, other nodes are all non-central nodes. (c) According to the LSI, nodes $1$ and $2$ are central nodes.}
\label{fig_Illustration}
\end{figure}

As shown in figure 1, we construct an illustration network by connecting two complete subnetworks through a single edge. Intuitively, this network has two communities (\{1,3,4,5\} and \{2,6,7\}) and nodes 1 and 2 should be treated as central nodes. However, neither GMD nor LMD can exactly identify those two central nodes. The top-$k$ highest-degree nodes may indeed belong to $\textless k$ communities and thus are not all central nodes, and the LMD can't identify central nodes with relatively smaller degrees.

Inspired by these shortcomings, we propose the so-called LSI index. The LSI of an arbitrary node $i$ reads

\begin{equation}
LSI_i = ({{k_i} - \frac{1}{{{k_i}}}\sum\limits_{j \in {\Gamma _i}} {{k_j}} })/ ({{k_i} + \frac{1}{{{k_i}}}\sum\limits_{j \in {\Gamma _i}} {{k_j}} }),
\end{equation}
where $k_i$ and $k_j$ represent the degree of node $i$ and node $j$, and $\Gamma_i$ denotes the set of $i$'s neighbors. Obviously, the value of LSI lies between -1 and 1, and the larger LSI corresponds to the higher local centrality. Without loss of generality, in this paper we treat nodes with $LSI_i \geq 0$ as central nodes for community detection, while other nodes are non-central nodes. As shown in Fig. 1(c), LSI identifies nodes $1$ and $2$ as central nodes.

We adopt the fitness function $F$ \cite{Lancichinetti2009,Zhou2007} as the quality function. For a community $C_i$, its $F$ value is

\begin{equation}
F(C_i) = \frac{d_{in}(C_i)}{d_{in}(C_i) + d_{out}(C_i)}, 
\end{equation}
where $d_{in}(C_i)$ is the internal degree (i.e., the twice number of edges in $C_i$), and $d_{out}(C_i)$ is the external degree (i.e., the number of edges connecting nodes in $C_i$ with nodes outside $C_i$). The fitness function $F$ for the whole network is defined as the sum of $F(C_i)$ of all communities, as
\begin{equation}
F = \sum_{i=1}^m{F(C_i)},
\end{equation}
where $m$ is the total number of communities.

Detailed procedures of the greedy algorithm are described below. Firstly, we identify all central nodes with LSI values no less than zero. Then we start the community expansion procedure. Initially, we choose one central node randomly as the seed of a provisional community $C_1$. At each step, we search all unassigned nodes that are neighboring to at least one node in $C_1$. Our goal is to find one node whose addition can maximize $F(C_1)$ and add this node to $C_1$. In case more than one node can maximize $F(C_1)$, we randomly choose one of them. We repeat this step until any addition of one neighbor can't increase $F(C_1)$. If $F(C_1)\leq 0.5$, the provisional community $C_1$ is disbanded and the current expansion becomes invalid, because the condition of weak community (i.e., $F(C_1)>0.5$) is not satisfied. At the same time, the central node is converted to an unassigned non-central node. We repeat this process until all provisional communities stop expanding. After the expansion, some nodes may still be unassigned and form one or more connected subgraphs. For each subgraph, we define its potential neighbors as all assigned nodes neighboring to at least one node of this subgraph. Then each subgraph is allotted to a provisional community who contains the potential neighbor with the highest LSI value. In case more than one provisional community satisfies such condition, we randomly choose one of them. Although the allocation of unassigned nodes may decrease the quality function, it is reasonable that the optimal community structure is not always with the largest quality function \cite{Zhang2014}. In the above procedure, the different order in choosing the central nodes may lead to different results, so we carry out the algorithm many times and adopt the result with the highest $F$.

\begin{table} 
\small
\centering  
\caption{Structural features of the six real networks under consideration. $N$, $L$ and $m$ represent the number of nodes, edges and communities respectively, and $\langle k\rangle$ represents the average degree. }
\begin{indented}
\item[]\begin{tabular}  
{>{\columncolor{white}}rccccc}  
\hline
\rowcolor[gray]{0.9}   Networks &$N$ &$L$   &$m$ &$\langle k \rangle$ &references \\ 
\hline
karate   &34   &78  &2  &4.5882 &\cite{Zachary1977} \\  
dolphins   &62 &159 &4   &5.129 &\cite{Lusseau2004} \\   
football   &115 &613    &12&10.6609 &\cite{Girvan2002} \\  
polblogs   &1222 &16714    &2&27.3552 &\cite{Lada2005} \\  
polbooks   &105 &441    &3&8.4 &\cite{Krebs0000},\cite{Newman0000} \\ 
highschool   &248 &1004    &8    &8.0968 &\cite{Moody2001} \\ 
\hline  
\end{tabular} 
\end{indented} 
\end{table} 

In this paper, we use two kinds of networks. One is the real networks  with known ground truth (i.e., recognized community structure), the other is the LFR benchmark networks generated artificially \cite{Lancichinetti2008}. We consider six real networks, with detailed description as follows. (i) Karate network \cite{Zachary1977}. A network of friendship in a karate club in an American university. The club split into two parts after a dispute between the coach and the treasurer. (ii) Dolphins network \cite{Lusseau2003}. This network contains 62 dolphins living in Doubtful Sound, New Zealand. An edge exists between two dolphins if they are observed together more often than expected by chance from 1994 to 2001. This network contains two communities \cite{Lusseau2004}. (iii) Football network \cite{Girvan2002}. A network of American football games between Division IA colleges during regular season Fall 2000. The nodes denote the 115 teams and the edges represent 613 games played in the course of the year. Each conference form a community and games are more frequent between members of the same conferences. (iv) Polblogs network \cite{Lada2005}. A directed network of hyperlinks between weblogs on US politics around the time of the 2004 presidential election. We convert this network to an undirected network by treating all directed links as undirected edges and reserve the largest component. Each weblog has a political leaning: blue (left or liberal) or red (right or conservative), and thus this network contains two communities. (v) Polbooks network \cite{Krebs0000,Newman0000}. In this network, nodes represent books about US politics sold in Amazon.com, and edges connecting frequently co-purchasing book pairs \cite{Lu2012}. Each book is assigned a political tag (liberal, neutral or conservative), and thus the network is consisted of three communities. (vi) Highschool network \cite{Moody2001}. A directed network represents the friendship choices made by students from different grades. We also convert this network to an undirected network. Fundamental structural statistics of these six networks are shown in Table 1.

\begin{figure}[H]
\setlength{\abovecaptionskip}{0.cm}
\setlength{\belowcaptionskip}{-0.cm}
\centering
	\includegraphics[width=0.8\textwidth]{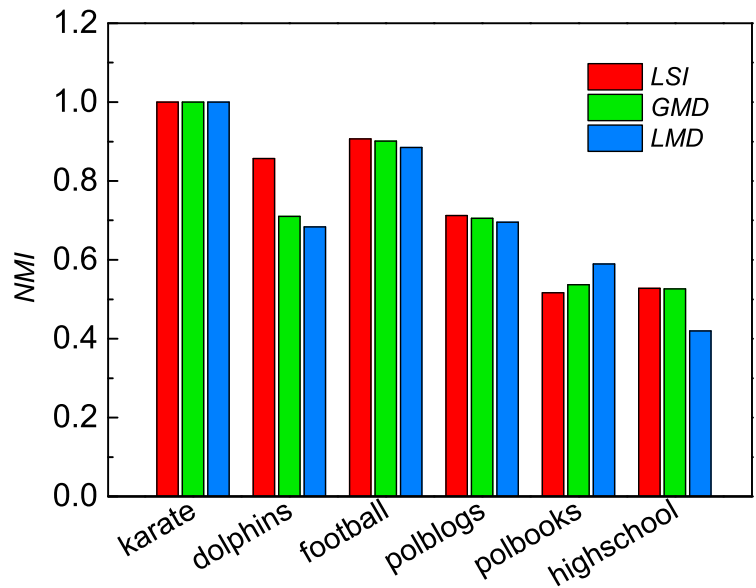}   
    \caption{ The NMI values of the three indices on the six real networks with known ground truth.  }
\label{fig_Illustration}
\end{figure}

LFR networks \cite{Lancichinetti2008} are classical benchmark networks generated by computer. In an LFR network, degrees are distributed in a power law with exponent $2<\gamma<3$, and the sizes of communities also follow a power-law distribution with exponent $1<\beta< 2$. Besides, the community size $s$ and node degree $k$ satisfy the constraints $s_{min}>k_{min}$ and $s_{max}>k_{max}$. An important mixing parameter $\mu$ represents the ratio between the external degree of an arbitrary node with respect to its community and the total degree of this node. With the increase of $\mu $, the community structure of the network becomes ambiguous.

\section{Results}

\begin{figure}[H]
\setlength{\abovecaptionskip}{0.cm}
\setlength{\belowcaptionskip}{-0.cm}
\centering
	\includegraphics[width=0.8\textwidth]{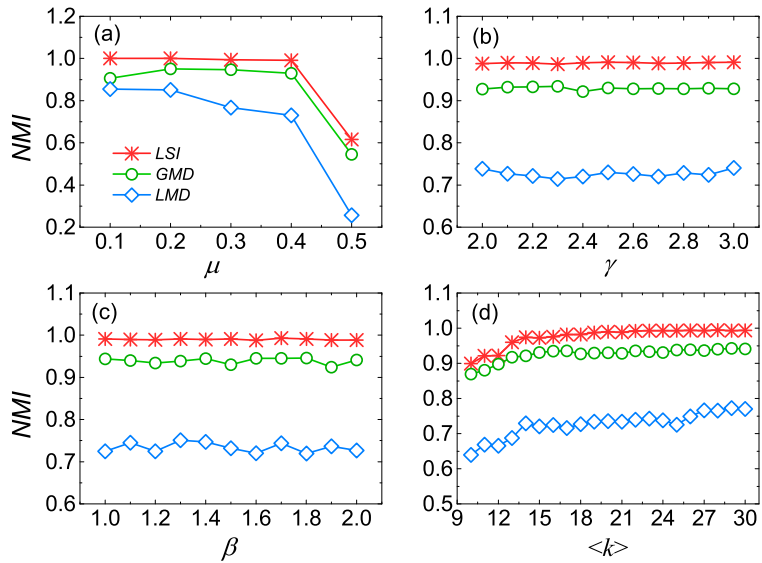}  
    \caption{How NMI changes with varying parameters in LFR networks. The default values of the four parameters are $\mu=0.4$, $\gamma=2.5$, $\beta=1.5$ and $\langle k\rangle=20$. In each plot, only one parameter is varying. The results are averaged over 10 networks with a fixed number of nodes $N=1000$. 
 } 
\label{fig_Illustration}
\end{figure}

We evaluate algorithms' performance by NMI \cite{Kuncheva2004}: the higher the NMI is, the better the algorithm is. Figure 2 shows that LSI can obtain better community structure than LMD and GMD for the real networks except the polbooks network.  Figure 3 compares the performance of the three indices on LFR networks. It is shown that LSI remarkably outperforms GMD and LMD. 

Figure 4 shows the relationship between nodes’ LSI values and degrees for a real network and an artificially generated network. The difference between LSI and degree is remarkable, namely LSI can identify some low-degree nodes as central nodes while reject some high-degree nodes. If one would like to correctly detect a real community, it is natural to expect that one or more nodes in this community should be first identified as central nodes. If none of nodes in this community are central nodes, the probability that this community could be perfectly detected is very tiny. We set the number of central nodes identified by GMD being equal to the number of central nodes identified by LSI, and then compare how many real communities having central nodes by the three indices. As shown in Tables 2 and 3, more real communities have at least one central node by LSI than by LMD or GMD. In particular, as shown in Table 3, for LFR benchmark networks, most communities have at least one central node according to LSI, however, more than a half communities do not have any central nodes by GMD and LMD. Indeed, LMD only identify very few central nodes and most communities do not have any central nodes accordingly. This may be the reason why LSI can outperform LMD and GMD in community detection.

\begin{figure}[H]
\setlength{\abovecaptionskip}{0.cm}
\setlength{\belowcaptionskip}{-0.cm}
\centering
	\includegraphics[width=0.8\textwidth]{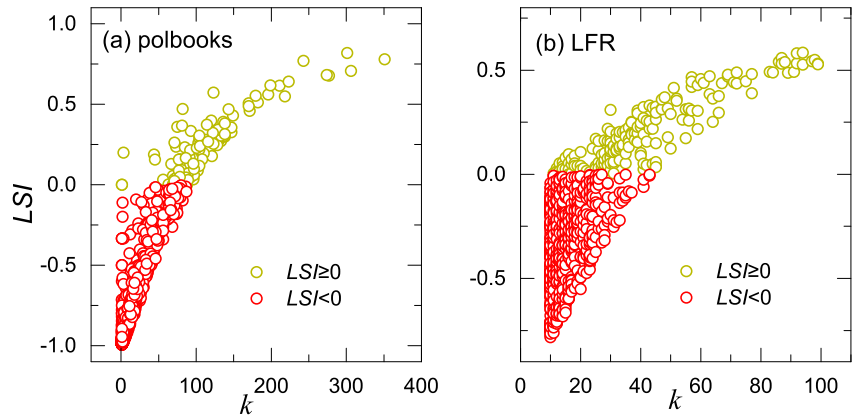}   
    \caption{LSI versus degree. Each data point represents one node. Nodes with  $LSI\geq0$ are colored by yellow, while nodes with  $LSI<0$ are colored by red. (a) The result for the polblogs network. (b) The result for the LFR network with parameters  $N=1000$, $\langle k\rangle=20$, $k_{max}=100$, $\gamma=2.5$, $\beta=1.5$ and $\mu=0.2$, where $k_{max}$ is the maximal degree. } 
\label{fig_Illustration}
\end{figure}

\begin{table}
\centering  
\caption{The number of real communities $m$ and the number of central nodes identified by the three indices on the six real networks. Each number in bracket is the number of real communities containing at least one central node. The number of central nodes for GMD is set to be equal to the number of central nodes identified by LSI.  }
\begin{indented}
\item[]\begin{tabular}
{>{\columncolor{white}}ccccccc}  
\hline
\rowcolor[gray]{0.9}   indicators &karate &dolphins &football &polblogs &polbooks   &highshool\\ 
\hline
$m$  &2 &4  &12   &2  &3  &8   \\   
LSI          &5(2)   &19(4)  &65(12)  &123(2) &24(3)  &80(7)  \\  
LMD          &2(2)   &5(3)   &32(11)  &6(2)   &3(3)   &12(6)   \\  
GMD          &5(2)   &19(3)  &65(12)  &123(2) &24(3)  &80(7) \\ 
\hline
\end{tabular}
\end{indented}  
\end{table} 

\begin{table}  
\centering  
\caption{The number of real communities $m$ and the number of central nodes identified by the three indices on the LFR networks. Each number in bracket is the number of real communities containing at least one central node. The number of central nodes for GMD is set to be equal to the number of central nodes identified by LSI.}
\begin{indented}
\item[]\begin{tabular}  
{>{\columncolor{white}}ccccccccc}  
\hline
\rowcolor[gray]{0.9}   indicators &$\mu=0.1$ &$\mu=0.2$ &$\mu=0.3$ &$\mu=0.4$ &$\mu=0.5$   \\  
\hline
$m$  &31 &31  &30   &35  &32   \\    
LSI          &242(29)   &200(29)  &187(25)  &176(27) &166(23)   \\ 
LMD          &9(8)   &6(4)   &5(3)  &5(4)   &4(4)    \\   
GMD      &242(13)    &200(15)  &187(15)  &176(15) &166(15)   \\  

\hline
\end{tabular}  
\end{indented}
\end{table}

\begin{figure}[H]
\setlength{\abovecaptionskip}{0.cm}
\setlength{\belowcaptionskip}{-0.cm}
\centering
	\includegraphics[width=0.8\textwidth]{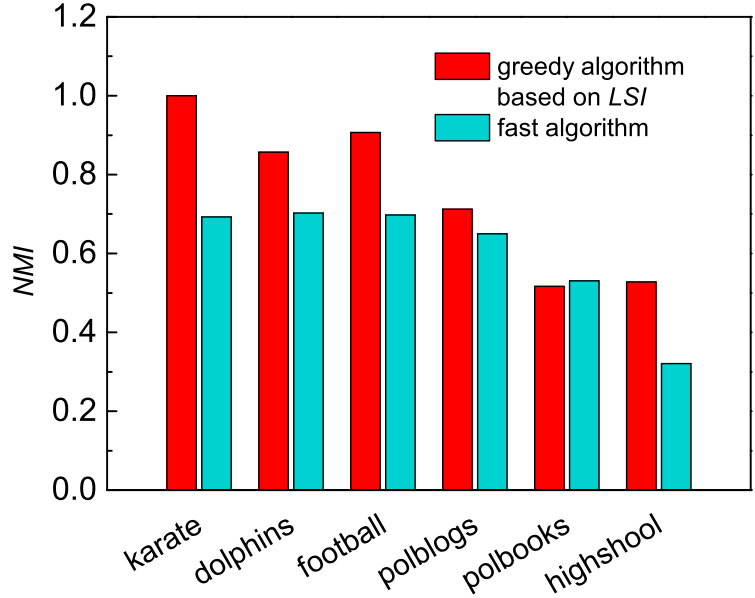}   
    \caption{The NMI values of the proposed greedy algorithm based on LSI and the Newman fast algorithm for the six real networks. }
\label{fig_IIlustration}
\end{figure}

\begin{figure}[H]
\setlength{\abovecaptionskip}{0.cm}
\setlength{\belowcaptionskip}{-0.cm}
\centering
	\includegraphics[width=0.8\textwidth]{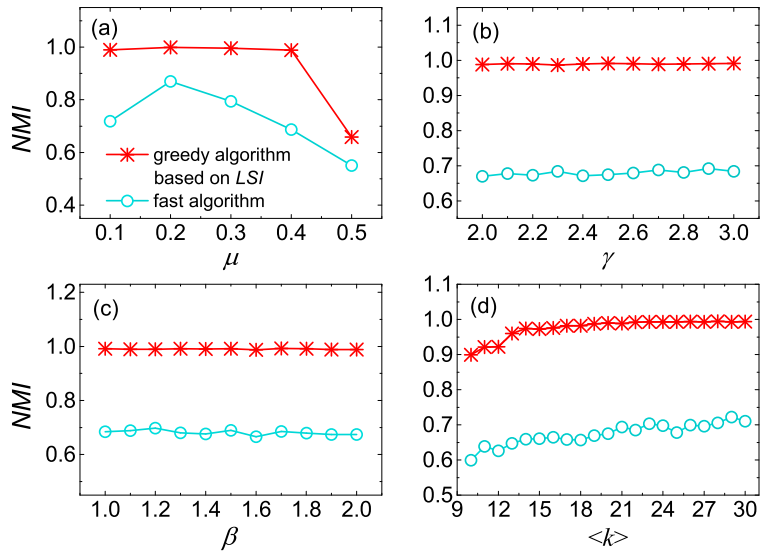}    
    \caption{The NMI values of the proposed greedy algorithm based on LSI and the Newman fast algorithm for the LFR networks. The default values of the four parameters are $\mu=0.4$, $\gamma=2.5$, $\beta=1.5$ and $\langle k\rangle=20$. In each plot, only one parameter is varying. The results are averaged over 10 networks with a fixed number of nodes $N=1000$.   }
\label{fig_IIlustration}
\end{figure}

The Newman fast algorithm is the most well-known greedy algorithm. It starts with every node being a single community, and then join a pair of provisional communities at each step to maximize the modularity $Q$, until all nodes agglomerated together to form one big community. The progress can be represented as a dendrogram, namely a tree that shows the order of all joins. The best community partition is obtained by cutting the dendrogram at the layer corresponding to the maximal $Q$. We compare the Newman fast algorithm with the proposed greedy algorithm based on LSI. As shown in figure 5, the proposed algorithm performs better than the Newman fast algorithm on real networks except the polbooks network. Figure 6 reports the comparison on LFR networks, indicating that the proposed algorithm outperforms the Newman fast algorithm and the superiority is robust.

\section{Conclusion and Discussion}

In this paper, we proposed a novel index LSI and a variant of greedy expanding algorithm that ensures all provisional communities are weak communities (i.e., internal degree is larger than external degree). The LSI performs remarkably better than the GMD and LMD in detecting communities if we apply those indices to identify central nodes for community expanding. In addition, the proposed greedy algorithm based on LSI outperforms the well-known Newman fast algorithm. 

\begin{figure}[H]
\setlength{\abovecaptionskip}{0.cm}
\setlength{\belowcaptionskip}{-0.cm}
\centering
	\includegraphics[width=0.8\textwidth]{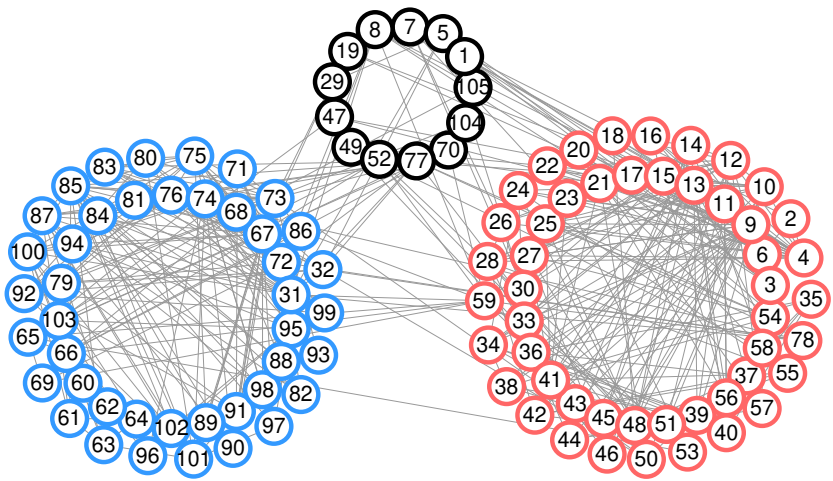}   
    \caption{Illustration of the community structure of the polbooks network. Red, blue and black nodes belong to conservative, liberal, and neutral communities, respectively.} 
\label{fig_Illustration}
\end{figure}

Polbooks is the only network where the proposed algorithm is slightly worse than benchmark algorithms. Figure 7 illustrates the community structure of the polbooks network, from which we can observe that the neutral community is not a weak community, with external edges far more than internal edges. As our algorithm avoids weak communities, it may not perform well if some ground-truth communities themselves are not weak communities. It is hard to say whether this is a disadvantage or an advantage, since if a ground-truth community is not a weak community, it usually contains some domain-specific reasons and thus may not be a typical community from the purely topological viewpoint.   

The proposed method can be extended to handle directed networks and weighted networks by slightly modifying the definition of LSI. If we allow non-central nodes to be merged to more than one community, the proposed algorithm can deal with the overlapping community structure. We can also add a free parameter $\alpha$ to the quality function as $F(C_i) = {d_{in}(C_i)}/{[d_{in}(C_i) + d_{out}(C_i)]^\alpha}$, and then uncover the hierarchical community structure with different resolutions via tuning the parameter $\alpha$.

\section*{Acknowledgements}
\addcontentsline{toc}{section}{Acknowledgements}
This work was supported by the National Natural Science Foundation of China under Grant Nos. 71874172, 11975071, and the Thousand Talents Program of Sichuan Province under Grant No. 17QR003.

\end{document}